\documentclass[prl,aps,amsmath,amssymb,showpacs,twocolumn]{revtex4-1}

\usepackage{bm,color,float,dcolumn,amssymb,graphicx,hyperref,subfigure}

\begin{document}

\title{Possible SU(3) Chiral Spin Liquid on the Kagome Lattice}

\author{Ying-Hai Wu}
\email{yinghai.wu@mpq.mpg.de}

\author{Hong-Hao Tu}
\email{hong-hao.tu@mpq.mpg.de}

\affiliation{Max-Planck-Institut f{\"u}r Quantenoptik, Hans-Kopfermann-Stra{\ss}e 1, 85748 Garching, Germany}

\date{\today}

\begin{abstract}
We propose an SU(3) symmetric Hamiltonian with short-range interactions on the Kagome lattice and show that it hosts an Abelian chiral spin liquid (CSL) state. We provide numerical evidence based on exact diagonalization to show that this CSL state is stabilized in an extended region of the parameter space and can be viewed as a lattice version of the Halperin 221 fractional quantum Hall (FQH) state of two-component bosons. We also construct a parton wave function for this CSL state and demonstrate that its variational energies are in good agreement with exact diagonalization results. The parton description further supports that the CSL is characterized by a chiral edge conformal field theory (CFT) of the SU(3)$_1$ Wess-Zumino-Witten type.
\end{abstract}
\maketitle

{\em Introduction} -- Topological aspects of condensed matter have been actively studied since the discovery of the quantum Hall effect. An important development in this area is the concept of topological order~\cite{wen2004}, which describes phases that can not be distinguished using the conventional symmetry breaking paradigm but exhibit exotic topological properties such as fractional charge, anyonic braiding statistics, and ground state degeneracy on high genus manifold. Besides the FQH states, it has long been speculated that there could be topologically ordered states in antiferromagnetic spin systems (called spin liquid) which do not break the crystalline symmetries or spin rotation symmetries. To suppress the tendency of magnetic ordering, one should consider frustrated lattices in which no simple alignment of spins can achieve the lowest energy under antiferromagnetic exchange. The Kagome lattice has been very promising in this regard, and recent numerical and experimental studies indeed point to the existence of spin liquid states in certain systems~\cite{yan2011,han2012,jiang2012,depenbrock2012,he2014,gong2014,bauer2014,hu2015}.

The theoretical study of strongly correlated spin systems is generally very difficult. Being motivated by the large $N$ expansion in gauge field theory, it has been proposed that one may investigate SU(N) spin systems using similar perturbative methods (organized in powers of $1/N$) to obtain some hints about the physics of SU(2) spins~\cite{arovas1988,affleck1988b,read1989,hermele2009}. One may worry that the perturbative results obtained in the large $N$ limit would not be applicable when $N$ is small, so other theoretical methods and numerical calculations are also essential in understanding the physics. Exactly solvable models, such as the Uimin-Lai-Sutherland model~\cite{uimin1970,lai1974,sutherland1975}, the Haldane-Shastry type models~\cite{haldane1988,shastry1988,kawakami1992,ha1992}, the Affleck-Kennedy-Lieb-Tasaki type models~\cite{affleck1987,affleck1991,chen2005,greiter2007a,greiter2007b,arovas2008}, have been designed and they provide useful insight into SU(N) spin systems. Another widely used method is to decompose the spins as bosonic or fermionic partons and build exotic spin states using mean field parton states supplemented with Gutzwiller projection. Based on different approaches, a rich variety of physical phenomena has been revealed in SU(N) spin systems~\cite{azaria1999,zhang2001,harada2003,corboz2011,bieri2012,honerkamp2004,cazalilla2009,gorshkov2010,manmana2011,nonne2013}.

It might appear at first sight that SU(N) spin systems are not relevant in experiments because the spins in solid state systems are almost all due to electrons so belong to the SU(2) group. However, it was proposed~\cite{li1998,yamashita1998} that the SU(4) Heisenberg model might describe certain materials in which SU(4) symmetry arises from coupled spin and orbital degrees of freedom~\cite{kugel1973}. There has also been substantial progress in experiments using cold atoms with several internal states, which brings SU(N) spin systems even closer to experimental reality~\cite{taie2010,taie2012,zhang2014,scazza2014}. The atomic species, lattice configurations, and the forms of interactions in cold atom experiments can be tuned in a wide range~\cite{bloch2008,cazalilla2014}, which would enable us to explore the rich physics associated with SU(N) spins.

The connection between FQH and CSL states was revealed in a seminal work by Kalmeyer and Laughlin~\cite{kalmeyer1987}, which demonstrated that bosonic FQH states can be mapped to spin states. In this example, one maps the spin-$1/2$ degree of freedom on a lattice site to a boson and impose the hard-core constraint such that there is at most one boson per site. This can be generalized to cases where the lattice sites have higher spins of the SU(2) group \cite{greiter2009}. We will explain below how to map SU(3) spins to bosons and establish a correspondence between SU(3) CSL and FQH states of two-component bosons.

Being equipped with the mapping between FQH and CSL states, we can use some techniques developed for FQH states to understand CSL. One fruitful way in the FQH context is to express FQH wave functions as chiral correlators of CFT~\cite{moore1991,nielsen2012}. The advantage of this approach is that, by using CFT null field technique~\cite{nielsen2011}, it can give a parent Hamiltonian with the CSL as its exact ground state~\cite{nielsen2012,tu2013a,tu2014b,bondesan2014} (see Refs.~\cite{schroeter2007,greiter2014} for alternative ways of deriving parent Hamiltonians). However, these parent Hamiltonians usually contain long-range interactions. To be more realistic, it is of great importance to test whether the CSL states constructed from CFT can be stabilized using Hamiltonians involving only simple short-range interactions~\cite{nielsen2013,glasser2015}.

{\em Mapping SU(3) Spins to Bosons} --- To make connections between SU(3) spin models and FQH states of two-component bosons, we briefly review their properties. The generators of the SU(3) group are usually chosen to be the eight $3\times3$ Gell-Mann matrices $\lambda_i$ ($i=1,2,\cdots,8$). For a lattice in which each site is described by the fundamental representation and the whole system is described by an SU(3) invariant Hamiltonian, the local Hilbert space dimension is three and there are two U(1) symmetries. To formulate a boson description, we may interpret the lattice as being occupied by two-component bosons (the two internal states are labeled as $\uparrow$ and $\downarrow$). Imposing the hard-core constraint that allows for at most one boson on each site results in a local Hilbert space dimension three ({\em i.e.} empty, one $\uparrow$ boson, and one $\downarrow$ boson). The two U(1) symmetries correspond to the particle number conservations of these two types of bosons.

The simplest FQH state of two-component bosons is the Halperin 221 state at filling factor $2/3$~\cite{halperin1983}
\begin{eqnarray}
\Psi_{221} = \prod^{M}_{s>t=1} (z^{\uparrow}_s-z^{\uparrow}_t)^2 (z^{\downarrow}_s-z^{\downarrow}_t)^2 \prod^{M}_{s,t=1} (z^{\uparrow}_s-z^{\downarrow}_t),
\end{eqnarray}
where $z=x+iy$ is the complex coordinate in two dimensions and the superscripts indicate the internal states. The low-energy properties of this state is encoded in the Chern-Simons theory with the Lagrangian density
\begin{eqnarray}
{\mathcal L} = \frac{1}{4\pi} K_{IJ} \epsilon^{\mu\nu\rho} a^{I}_{\mu} \partial_\nu a^{J}_{\rho},
\end{eqnarray}
where $K_{IJ}$ is the $2\times2$ matrix
\begin{eqnarray}
\left(\begin{array}{cc}
2 & 1 \\
1 & 2
\end{array}\right)
\label{Kmatrix}
\end{eqnarray}
One characteristic signature of topologically ordered states, the ground state degeneracy on torus, can be deduced from this Chern-Simons theory as $|{\rm det}K|=3$. The Chern-Simons action also provides useful information about its edge physics: the $K$ matrix has two positive eigenvalues, so there are two copropagating edge modes described by $U(1){\times}U(1)$ bosons. For bosons in the lowest Landau level, this state is the exact zero energy ground state if there are only contact interactions $\sum_{\sigma\tau} \delta({\mathbf r}^\sigma-{\mathbf r}^\tau)$ between the bosons regardless of their spins. The contact interaction forbids two bosons to appear at the same position, which is somewhat equivalent to the constraint of having at most one boson per site in the bosonic description of SU(3) spin models. In general, the spin models defined on a lattice appear to be very different from the simple continuum model, but their low-energy effective theories have the same action. This can be seen from the parton construction of the SU(3) CSL state (see below).

\begin{figure}
\includegraphics[width=0.49\textwidth]{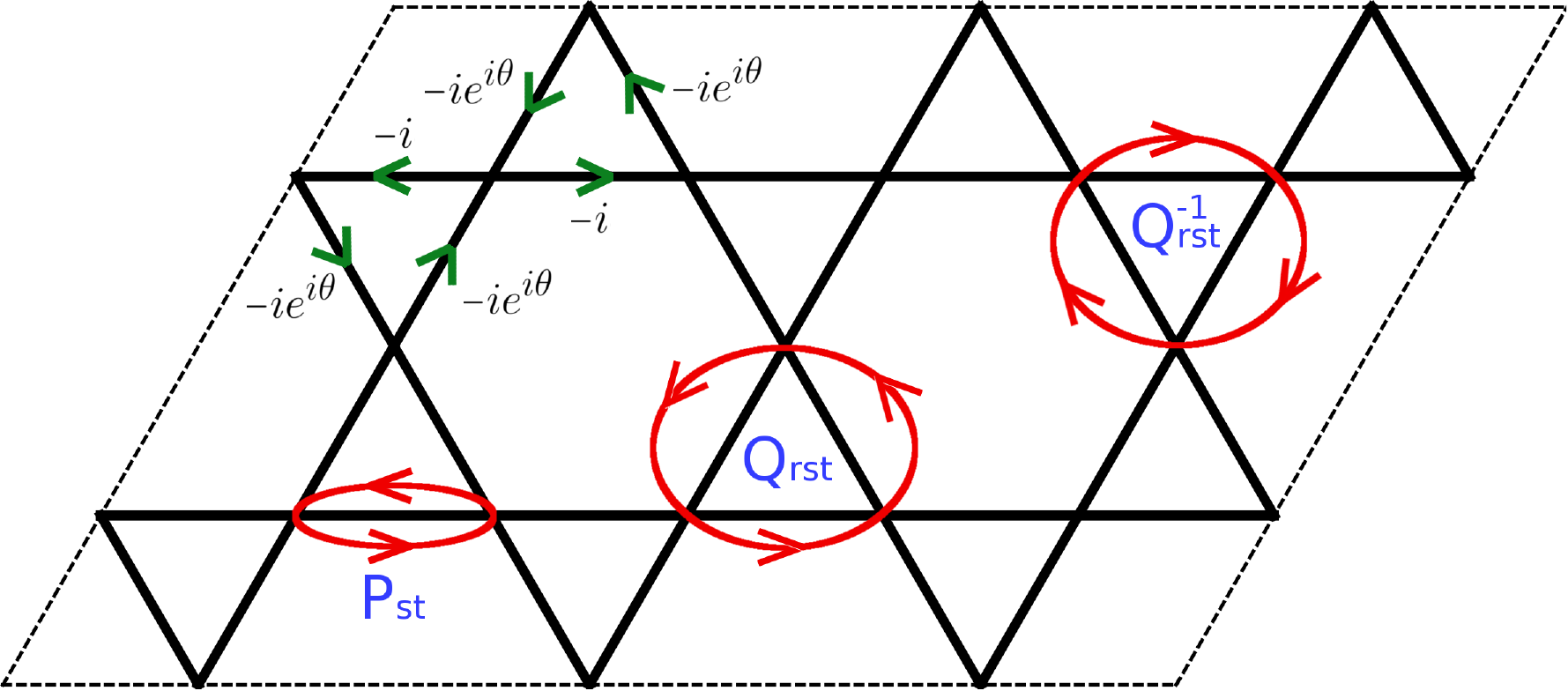}
\caption{The Kagome lattice with 18 sites. The red circles illustrate the three types of terms $P_{st}$, $Q_{rst}$, and $Q^{-1}_{rst}$ in the SU(3) Hamiltonian~(\ref{eq:Hamiltonian}). The green arrows on the small triangles and the numbers in their vicinity give the hopping phases in the parton mean field Hamiltonian~(\ref{eq:partonMF}).}
\label{Figure1}
\end{figure}

\begin{figure}
\includegraphics[width=0.45\textwidth]{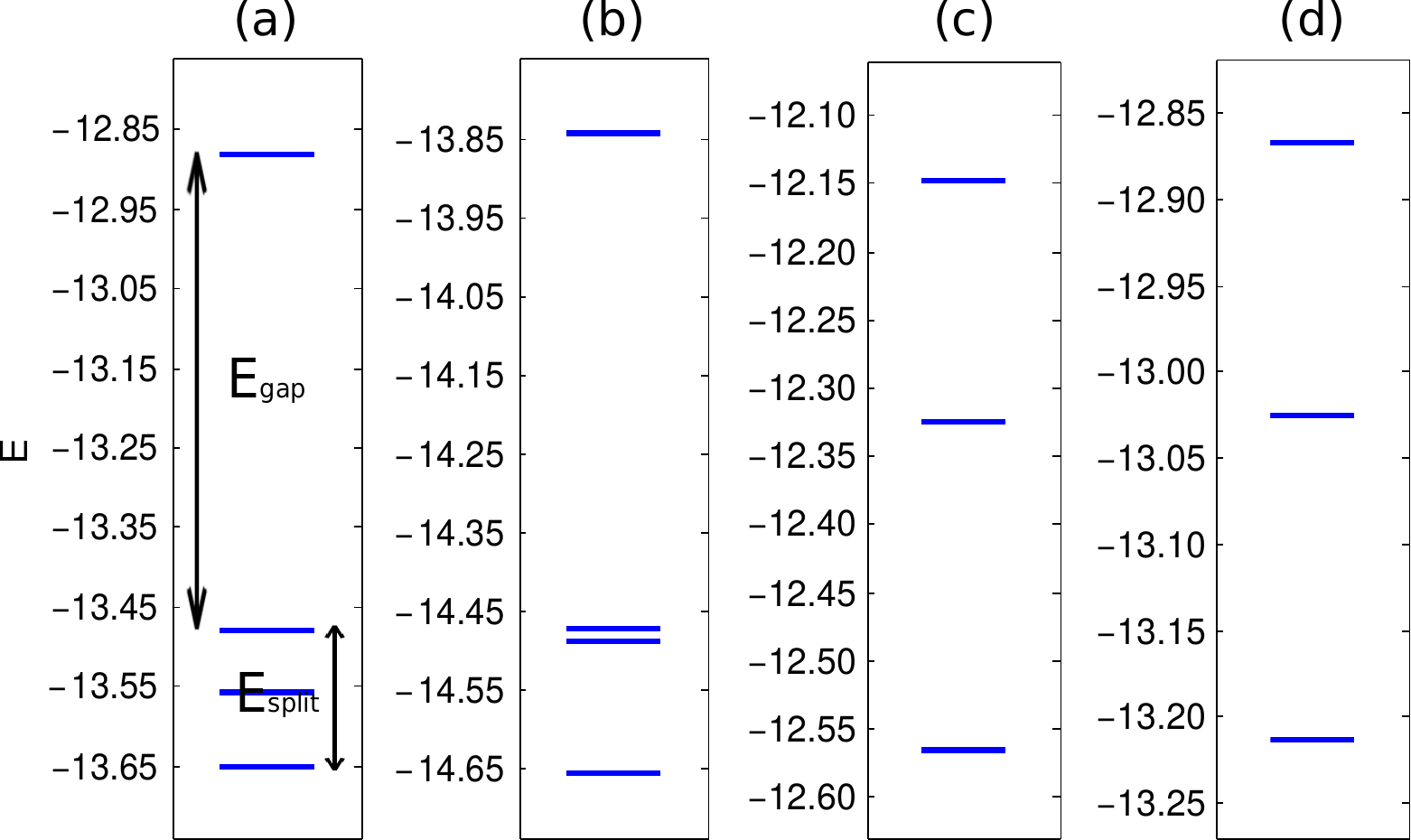}
\caption{(Color online) Energy spectra on the Kagome lattice with 18 sites. (a) $K_1=0.6$ and $K_2=0.4$ with PBC; (b) $K_1=0.6$ and $K_2=0.5$ with PBC; (c) $K_1=0.6$ and $K_2=0.4$ with OBC; (d) $K_1=0.6$ and $K_2=0.5$ with OBC. The value of $J$ is fixed at 1 in all calculations.}
\label{Figure2}
\end{figure}

\begin{figure}
\includegraphics[width=0.49\textwidth]{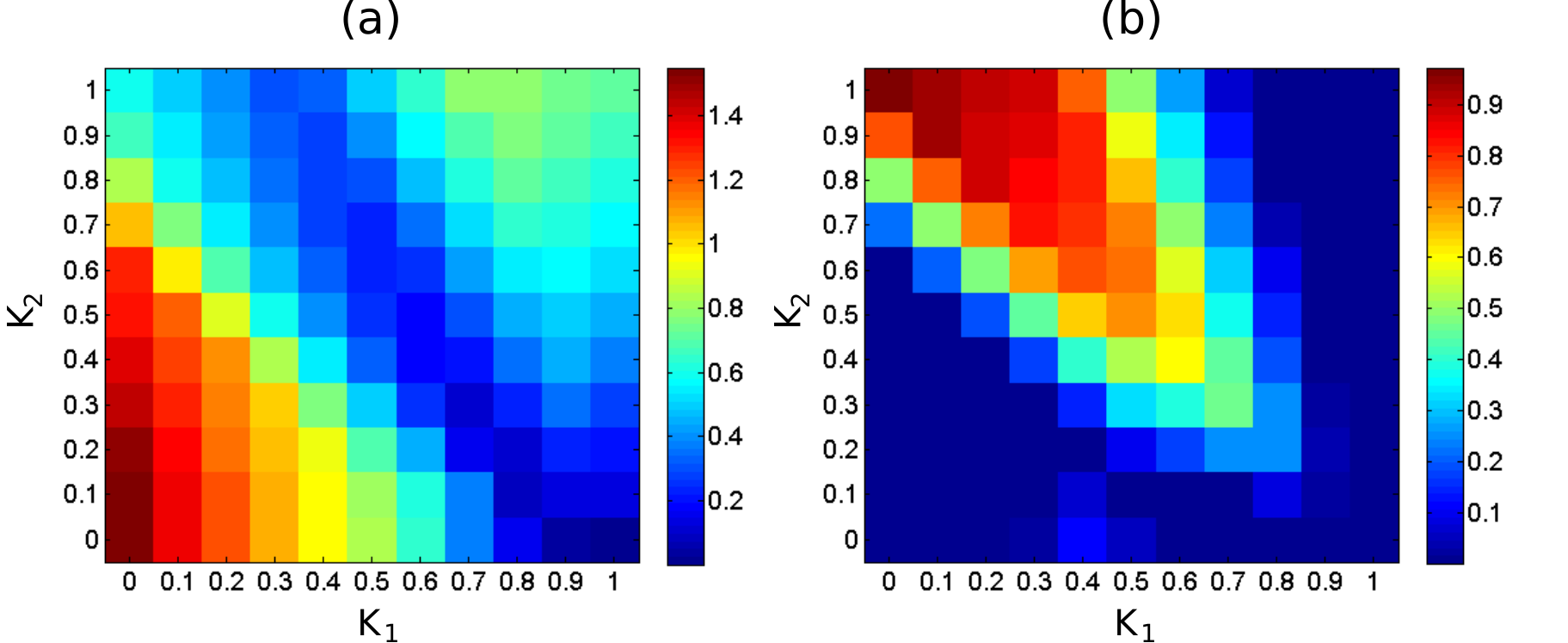}
\caption{(Color online) $E_{\rm split}$ and $E_{\rm gap}$ on the Kagome lattice with 18 sites. (a) $E_{\rm split}$ at $K_1=0.0,0.1,\cdots,1.0$ and $K_2=0.0,0.1,\cdots,1.0$; (b) $E_{\rm gap}$ at $K_1=0.0,0.1,\cdots,1.0$ and $K_2=0.0,0.1,\cdots,1.0$. The value of $J$ is fixed at 1 in all calculations.}
\label{Figure3}
\end{figure}

{\em Exact Diagonalization} --- The CFT construction provides us parent Hamiltonians for which the SU(3) CSL states are exact ground states~\cite{tu2014b,bondesan2014}. These Hamiltonians inevitably contain long-range terms but they provide useful hints about what kind of short-range Hamiltonians might have ground states in the same phase. A general Hamiltonian can be written in terms of the Gell-Mann matrices, but SU(3) invariance imposes stringent constrains on the Hamiltonians and it is usually more convenient to express them in terms of swapping operators. For our purpose, we need to define two-body and three-body swapping operators $P_{st}$ and $Q_{rst}$. When $P_{st}$ is applied on a state, the spin states on the lattice sites $s$ and $t$ are exchanged. When $Q_{rst}$ is applied on a state, the spin states on lattice $r$, $s$ and $t$ are cyclically permuted in a counterclockwise way.

The short-range Hamiltonian we have studied is defined on the Kagome lattice with two-body terms acting on all nearest neighbors and three-body terms acting on all small triangles
\begin{eqnarray}
H &=& J \sum_{\langle{st}\rangle} P_{st} + (K_1 - i K_2) \sum_{\langle{rst}\rangle} Q_{rst} \nonumber \\
  &+& (K_1 + i K_2) \sum_{\langle{rst}\rangle} Q^{-1}_{rst},
  \label{eq:Hamiltonian}
\end{eqnarray}
where $Q^{-1}_{rst}$ means permuting the spin states clockwisely (equivalent to two counterclockwise permutations). In Fig.~\ref{Figure1}, we show a Kagome lattice with 18 sites ($3$ unit cells along one direction and $2$ unit cells along the other direction) and illustrate the terms in the Hamiltonian. The numerical results presented below are for this lattice but we have obtained similar results for the Kagome lattice with $12$ sites ($2$ unit cells along both directions). The Hamiltonian~(\ref{eq:Hamiltonian}) is SU(3) invariant, so its eigenstates belong to definite representations of the SU(3) group. We choose $J=1$ as the energy scale and vary $K_{1,2}$ over a broad range to search for the optimal values that may stabilize an SU(3) CSL corresponding to the Halperin 221 state.

The energy spectra for a few systems with periodic boundary condition (PBC) or open boundary condition (OBC) are shown in Fig.~\ref{Figure2}. For a system with PBC, the SU(3) CSL that we seek has three degenerate ground states in the thermodynamic limit but the ground states generally split in finite size systems. On the contrary, such a system has only one ground state if it has OBC. In both cases, the ground state(s) are separated from the excited states by an energy gap. The numerical results in Fig.~\ref{Figure2} are consistent with these theoretical expectations. We have also confirmed by explicit calculations that the ground states are SU(3) singlets. For the cases with PBC, the eigenstates also have good momentum quantum numbers and we found that the three quasi-degenerate ground states all have $K_x=0$ and $K_y=0$. The energy spectra on torus can be characterized quantitatively using two variables $E_{\rm gap}$ and $E_{\rm split}$ as shown in Fig.~\ref{Figure2}: the former is the difference between the third state and the fourth state and the latter is the splitting of the lowest three states. It is desirable to have a sufficiently large $E_{\rm gap}$ and a small enough $E_{\rm split}$. These two variables are plotted in Fig.~\ref{Figure3} for a wide range of parameters and one can see that such requirements are satisfied in a region around $K_1\approx0.6$ and $K_2\approx0.45$.

\emph{Parton Wave Functions} --- To gain further insights into the nature of the ground states of the SU(3) Hamiltonian (\ref{eq:Hamiltonian}), we now resort to a parton wave function description of the numerically observed CSL phase. This relies on a fermionic representation of the SU(3) spins, where the three local states are encoded using singly occupied fermions, $|\alpha\rangle =c^{\dagger}_{\alpha}|0\rangle$ ($\alpha=1,2,3$). The redundant states in the fermionic Hilbert space are removed by a Gutzwiller projector $P_{\mathrm{G}}$ which locally enforces single occupancy on each site, {\em i.e.} $\sum_{\alpha}c^{\dagger}_{s\alpha}c_{s\alpha}=1$ $\forall s$.

We assume that the partons are described by the free fermion Hamiltonian
\begin{equation}
H_{\rm parton} = \sum_{\alpha} \sum_{\langle{st}\rangle} f_{st} c^{\dagger}_{s\alpha} c_{t\alpha},
\label{eq:partonMF}
\end{equation}
where $f_{st}$ is the hopping parameter of fermionic partons between nearest neighbors (to be determined below). This Hamiltonian can be viewed as a ``mean field" theory of the original SU(3) spin problem. However, at the mean field level, the particle number constraint is only satisfied on average. A trial wave function in the physical spin Hilbert space should satisfy the single occupancy constraint rigorously, which can be obtained using Gutzwiller projection as
\begin{equation}
|\Psi \rangle =P_{\rm G}|\Psi _{\rm parton}\rangle,
\label{eq:partonWF}
\end{equation}
where $|\Psi _{\rm parton}\rangle$ is the Fermi sea ground state of~(\ref{eq:partonMF}) at 1/3 filling.

For our purpose of describing the numerically observed CSL state, we choose the hopping integral $f_{st}$ in~(\ref{eq:partonMF}) to be complex numbers whose phases depend on only one parameter $\theta$ as shown in Fig.~\ref{Figure1}. The value of $\theta$ determines the fluxes in the triangles and hexagons of the Kagome lattice. With this prescription, the parton Hamiltonian~(\ref{eq:partonMF}) with PBC has three energy bands and, at 1/3 filling, the lowest band is completely filled by fermionic partons. For OBC, the parton wave function is similarly constructed by assuming an open boundary for the parton Hamiltonian (\ref{eq:partonMF}). For both PBC and OBC, the Gutzwiller wave functions~(\ref{eq:partonWF}) are SU(3) singlets.

\begin{figure}
\centering
\includegraphics[scale=0.48]{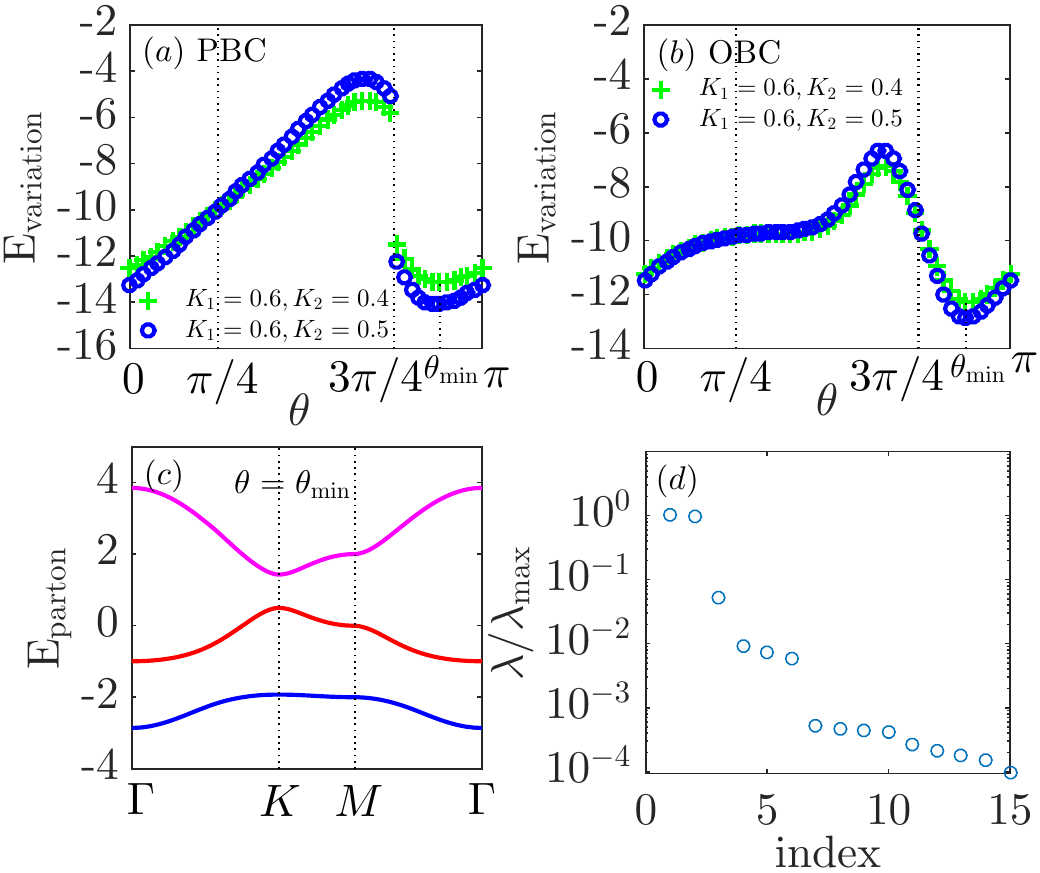}
\caption{(Color online) (a) PBC and (b) OBC variational energies of the parton trial states for the 18-site Kagome lattice as a function of the parameter $\theta$. The lowest variational energies for the Hamiltonian~(\ref{eq:Hamiltonian}) with $K_1=0.6$, $K_2=0.4$ (green crosses) and $K_1=0.6$, $K_2=0.5$ (blue open circles) both appear at $\theta_{\rm min}\approx0.88\pi$. For PBC, the gap between the lowest and the middle bands of the parton Hamiltonian vanishes at $\theta=\pi/4$ and $\theta=3\pi/4$ (denoted by two dotted lines). (c) Band structure of the parton Hamiltonian at the optimal variational point $\theta_{\rm min}$. The three bands are separated by energy gaps and have Chern numbers $-1$, $0$, and $+1$, respectively (from top to bottom). (d) (Normalized) eigenvalues of the overlap matrix of Gutzwiller wave functions with 15 different twisted boundary conditions for partons on the 48-site Kagome lattice. The existence of three large eigenvalues suggests that there are three linearly independent states on torus.}
\label{Figure4}
\end{figure}

To optimize the variational ansatz, we choose many different $\theta$ values and compute the energy of the wave function~(\ref{eq:partonWF}) with respect to the Hamiltonian (\ref{eq:Hamiltonian}) to select the one giving the lowest energy. This has been done in several cases on the lattice with 18 sites but we focus here on the following two sets of parameters, i) $K_{1}=0.6, K_{2}=0.4$ and ii) $K_{1}=0.6, K_{2}=0.5$. The variational energy as a function of $\theta$ is shown in Fig.~\ref{Figure4} (a) and (b). The best results in the two cases with PBC (OBC), which both appear at $\theta_{\rm min}{\approx}0.88\pi$, are $-13.12$ ($-12.30$) for $K_{2}=0.4$ and $-14.08$ ($-12.89$) for $K_{2}=0.5$. For PBC, they are quite close to the energies of the three quasi-degenerate ground states [see Fig.~\ref{Figure2} (a) and (b)]. The variational energies are, however, less satisfactory for OBC [see Fig.~\ref{Figure2} (c) and (d)].

For the optimal choice $\theta_{\rm min}$, the three parton energy bands of~(\ref{eq:partonMF}) have Chern numbers $-1$, $0$, $+1$, respectively [see Fig.~\ref{Figure4}(c) from top to bottom]. This means that the parton trial wave function (\ref{eq:partonWF}) describes a Gutzwiller projected Chern insulator with Chern number $+1$. To describe the three quasi-degenerate ground states on torus, one may construct parton wave functions by adopting twisted boundary conditions for the partons~\cite{zhang2012,tu2013b}. We have checked that, by computing the eigenvalues of the overlap matrix, 15 different twisted boundaries for partons on the 48-site Kagome lattice (4 unit cells in both directions) indeed yield three linearly independent states [see Fig.~\ref{Figure4}(d)]. Thus, these three parton wave functions provide a complete approximation of the ground state manifold on torus. Because the SU(3) CSL state may be interpreted as Gutzwiller projected Chern insulator with Chern number $+1$, one can proceed to derive its low-energy effective theory using functional path integral and the resulting theory turns out to be SU(3)$_{1}$ Chern-Simons theory~\cite{lu2014,liu2014}.

{\em Conclusion and Discussion} --- In this Rapid Communication, we have investigated an SU(3) symmetric Hamiltonian consists of short-range interactions on the Kagome lattice. Based on exact diagonalization results, we have identified an extended region in the parameter space where the system realizes an Abelian SU(3) CSL. A trial wave function constructed using fermionic partons and Gutzwiller projection is found to be a good approximation of the exact eigenstates. The parton description also helps us to deduce the low-energy effective theory. It has been shown in Ref.~\cite{sun2015} how to derive a Chern-Simons theory for SU(2) spin systems on arbitrary lattices, so one might expect that the Chern-Simons theory for the SU(3) CSL can also be derived without reference to partons.

In general, one can establish a mapping between SU(N) spins and (N-1)-component bosons, which suggests that SU(N) CSL and FQH states of multi-component bosons are closely related. It would be very interesting if one can also identify short-range interactions in other SU(N) spin systems that can host CSL states. Another exciting direction opened by our current work is to investigate SU(N) CSL with non-Abelian anyons. The CFT construction of multi-component non-Abelian FQH states~\cite{ardonne1999,ardonne2001} is a good starting point in this direction. The relation between these non-Abelian spin-singlet (NASS) states and the Halperin 221 state is very much the same as that between the Moore-Read state and the Laughlin state. While the Kalmeyer-Laughlin state is designed for spin-$1/2$ systems, the lattice version of the Moore-Read state may be realized in spin-$1$ systems. It should be possible to reformulate the NASS states in SU(N) spin systems where the spins are described by higher-dimensional representation of the SU(N) group. It would also require numerics to see if such states can be stabilized by sufficiently simple short-range Hamiltonians.

Upon finalizing the manuscript we noticed two recent preprints~\cite{nataf2016a,nataf2016b} on closely related topics.

{\em Acknowledgement} --- We are grateful to Meng Cheng for helpful discussions. HHT acknowledges A.~E.~B.~Nielsen and G.~Sierra for an earlier collaboration on the SU(N) CSL. Exact diagonalization calculations are performed using the DiagHam package for which we thank all the authors. This work is supported by the EU project SIQS.

\bibliography{Liquid}

\end{document}